\documentclass[aps, 10pt, prl,
               notitlepage, twocolumn, superscriptaddress,
               longbibliography,
               nofootinbib, floatfix,  secnumarabic,amssymb, 
               nobibnotes]{revtex4-2}
\usepackage{amsmath}
\usepackage{graphicx}
\usepackage{lipsum}  
\usepackage{xcolor}  
\usepackage{soul}
\usepackage{cancel}
\usepackage{amssymb}
\usepackage{multirow}
\usepackage[english]{babel}
\newcommand{\Cp}{c_{+}}
\newcommand{\Cm}{c_{-}}
\newcommand{\Cpm}{c_{\pm}}

\newcommand{\grad}{\bar{\nabla}}

\newcommand{\p}{\textbf{p}}

\begin{document}

\title{Phase separation by polar active transport}

\author{Sudipta Pattanayak}
\email{These authors contributed equally}
\affiliation{Collège de France, Université Paris Sciences et Lettres, Matière molle et biophysique, Paris 75231, France
}
\affiliation{Institut Curie, Université Paris Sciences et Lettres, Physique de la Cellule et Cancer, Paris Cedex 05 74248, France}
\author{Alfredo Sciortino}
\email{These authors contributed equally}
\affiliation{CytoMorpho Lab, Chimie Biologie Innovation, UMR8132, Université Paris Sciences et Lettres, Ecole Supérieure de Physique et Chimie Industrielles de la Ville de Paris, CEA, CNRS, Institut Pierre Gilles De Gennes, Paris 75005, France}
\author{Laurent Blanchoin}
\affiliation{CytoMorpho Lab, Laboratoire de Physiologie Cellulaire et Végétale, UMR5168, Université Grenoble-Alpes, CEA, INRA, CNRS, Interdisciplinary Research Institute of Grenoble, Grenoble 38054, France}
\affiliation{CytoMorpho Lab, Chimie Biologie Innovation, UMR8132, Université Paris Sciences et Lettres, Ecole Supérieure de Physique et Chimie Industrielles de la Ville de Paris, CEA, CNRS, Institut Pierre Gilles De Gennes, Paris 75005, France}
\author{Manuel Théry}
\affiliation{CytoMorpho Lab, Laboratoire de Physiologie Cellulaire et Végétale, UMR5168, Université Grenoble-Alpes, CEA, INRA, CNRS, Interdisciplinary Research Institute of Grenoble, Grenoble 38054, France}
\affiliation{CytoMorpho Lab, Chimie Biologie Innovation, UMR8132, Université Paris Sciences et Lettres, Ecole Supérieure de Physique et Chimie Industrielles de la Ville de Paris, CEA, CNRS, Institut Pierre Gilles De Gennes, Paris 75005, France}
\author{Jean-Fran\c cois Joanny}
\email{Corresponding, email: jean-francois.joanny@college-de-france.fr}
\affiliation{Collège de France, Université Paris Sciences et Lettres, Matière molle et biophysique, Paris 75231, France
}
\affiliation{Institut Curie, Université Paris Sciences et Lettres, Physique de la Cellule et Cancer, Paris Cedex 05 74248, France}

\begin{abstract}
We propose an active Cahn-Hilliard theory for the dynamics of a new type of phase transition where the driving force is not the direct interactions between the two separating components, but their active sorting by a third polar species. This third species transports the other two along its polarity in opposite directions, thus separating them. Inspired by recent experiments where molecular motors that walk in opposite directions along microtubules are sorted into separated domains, our theoretical description introduces a new mechanism for active phase separation and could serve as a model for the organization of biological material in space inside cells. We predict the formation of motor domains, and further show that they can either coarsen to form macroscopic phases or reach a finite micro- or mesoscopic steady state  size, these latter due to an arrest of coarsening through activity.

\end{abstract}

\maketitle

\section{Main Text}
In 1871, James Clerk Maxwell described how ``a being whose 
faculties are so sharpened that he can follow every molecule in its course''\cite[Chapter~22]
{maxwell_theory_1871} could achieve the unexpected result of sorting individual molecules based on their microscopic state,  potentially 
decreasing the system's entropy in the absence of external work. While we now know that this is impossible at equilibrium, even at the microscopic scale,  Maxwell's thought experiment insightfully highlights how, if 
one has 
access to information about the microscopic states of the system and can perform work at the microscopic scale, this work can be used to 
overcome locally the limitations of thermodynamics, at the cost of course of expending energy \cite{malgaretti_szilard_2022}.
 More than 150 year later, we do have a fuller understanding of how active systems can harness energy from 
their environment to 
perform tasks forbidden at 
equilibrium, a striking example being directed motion\cite{cates_active_2024,te_vrugt_metareview_2025}. One 
consequence is the ability of active particles to phase 
separate in space, using active processes to 
counteract the equilibrium's tendency towards a homogeneous state. 
\begin{figure}[t]
\centering
  \includegraphics[width=\columnwidth]{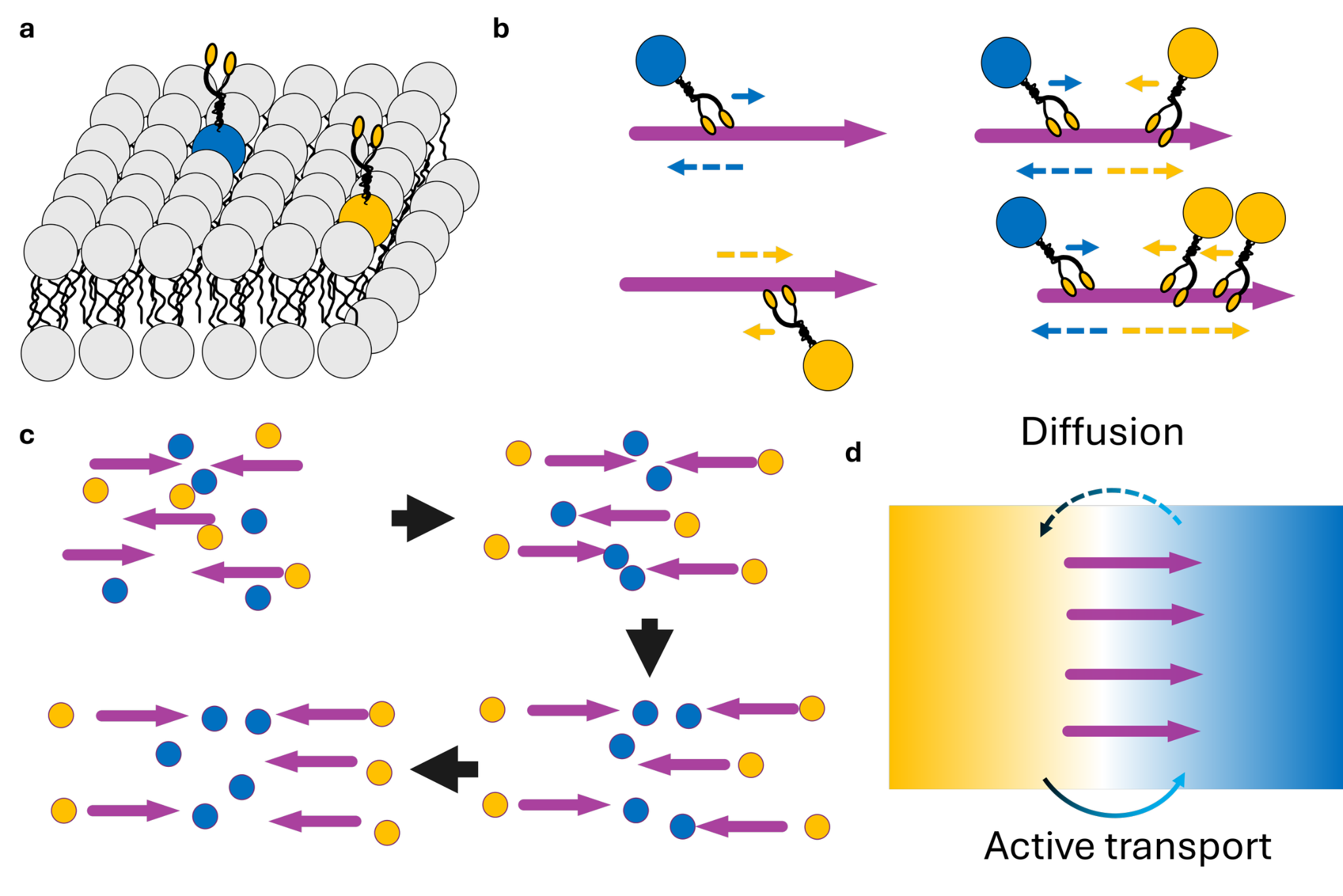}
  \caption{ {\bf a.} Schematics of the experimental system. Two types of motors, plus-end directed (blue) and minus end directed (yellow), are embedded in a fluid lipid membrane and diffuse freely. {\bf b.} While bound to lipids, motors move along  filaments (purple arrow) in either direction (solid blue/yellow arrow). At the same time, they exert a force on the filament in the opposite direction (dashed blue/yellow arrow). When both motors walk on a filament, their forces add up. {\bf c.} From an initially disordered state, motors first walk on filaments and separate in space. This creates a local force imbalance on filaments, that glide until the motor concentration at their side is such that no net force is exerted on the filaments. This results in the creation of motors-enriched domains separated by a filament polar interface. {\bf d.} When patterns have formed, filaments accumulate at the interface. Their pumping of motors  (active transport) is countered by passive diffusion.} 
  \label{fig:scheme}
\end{figure}
Active 
phase separation 
has been observed in a variety of systems, from living, such 
as bacteria and cells \cite{balasubramaniam_investigating_2021, liu_self-driven_2019, TakatoriPRL_2023}, to synthetic, such as self-propelled 
colloids\cite{van_der_linden_interrupted_2019} or reconstituted protein 
assemblies \cite{weyer_deciphering_2024, winter_phase_2024, yan_stochastic_2025, kant_bulk_2024}.
All these 
systems break detailed balance to establish 
energy fluxes that 
favor an otherwise forbidden spatial organization\cite{Marchetti_RMP_2013}. 
Theoretical studies of active phase separation, using models based on continuous fields,  have shown specific 
properties  that do not exist in passive phase separation, including reverse Ostwald 
ripening or robust formation of microdomains\cite{wittkowski_scalar_2014, 
cates_active_2024, tjhung_cluster_2018, Caballero_active_stress_2022, Saha_arrested_phaseseparation_2025}. Although the concept of active phase separation is very general only few quantitative experimental studies address its importance in biological systems. Yet in living cells, active transport by molecular motors along 
filaments of the cytoskeleton also offers possibilities to separate 
differentially matter in space. The inherent 
polarity of microtubules, allows for
motors to move in specific directions, towards their plus or minus end depending on the motor type. By this process, microtubules, segregate motors in space according to their directionalities.

In a way, the motor–microtubule complex thus functions as an ATP-powered microscopic "pump": by coupling motor directionality to motion along the polar microtubule lattice, plus- and minus-end-directed species are driven in opposite directions. This directionality-dependent sorting across a microtubule is a mechanism reminiscent of a Maxwell’s demon, albeit one being fueled by ATP hydrolysis so that no violation of the Second Law is taking place.

This sorting mechanism is used by cells \cite{demon1, demon2} to actively segregate and distribute 
cargoes in space in order to establish cell polarity or to form and maintain organelles. 
Several recent experiments aim at understanding the precise sorting using systems 
of filaments and molecular motors to reconstitute in vitro the self-organization of a polarized network, which drives the separation of motors in space  \cite{mitchison_self-organization_2021, Schaller2010, 
Sumino2012a, Huber2018a, Sciortino2021, memarian_active_2021}. Microtubules 
and kinesins, in particular, have allowed to build complex 
supramolecular assemblies sometimes resembling equilibrium structures 
such as foams or micelles but based on the formation of separated motor 
clusters along the filaments' polarity \cite{Lemma2021, 
de_luca_supramolecular_2024}.

Recently, we also introduced a system consisting of 
stabilized microtubules propelled by two opposite kinesin 
motors bound to a fluid membrane (Fig. \ref{fig:scheme}.a). 
We showed how motors of different types sort into domains 
separated by a polarized microtubule interface  \cite{utzschneider_force_2024}. This process is based on the ability of motors to both walk on the filaments and at the same time exert forces on them in opposite directions, until a balance is achieved (Fig. \ref{fig:scheme}.b-c).  The formation of motor domains and 
their dynamics strikingly resemble passive phase separation but is
microscopically fueled by the active "pumping" of motors. We also observed that domains grow only until a 
steady state is reached where both the force exerted by the motors on the 
microtubules and the total local motor fluxes vanish (Fig. \ref{fig:scheme}.d).  Here we show how this process (sketched in Fig~\ref{fig:scheme}) can be indeed considered as 
an example of active 
phase separation, whose driving force is
not the direct interactions between the two motor types but the 
bidirectional active transport of motors along microtubules.

By writing a Cahn-Hilliard theory for motors/microtubules mixtures, we find that under certain conditions, activity 
can slow
down the domain coarsening, giving rise to a stable steady-state formed by alternating regions of the two motor types or droplets of one motor in a continuous phase of the other one. 

In our model, the motors and microtubules are coarsened into 
density fields and the orientation of the microtubules 
into a  polarity vector field $\bf p$ defined by  local averaging of microtubule orientation. 

The three conserved fields are the two motors ($c_\pm$) and microtubule ($\phi$) concentrations. We schematize filaments as rectangles of lengths $L$ and $2d\ll L$, and the motors as circles of radius $d$. We define the corresponding  
area fractions as {$\hat{c}_\pm= \Cpm /\rho_0$} and $\psi=2Ld\phi$, 
where $\rho_0=1/\pi d^2$, and the total area fraction as $\hat{c}_{tot}=\hat{c}_+ +\hat{c}_ - +\psi$ 

The dynamics of the phase separation is 
obtained from the coupled dynamic equations of the concentrations and 
polarity fields \cite{bray_theory_2002}.
The system's behavior is described in terms of currents of the three 
components. We decompose here the currents into passive currents, driven 
by gradients in chemical potentials and active currents, due to 
the motor motion with speed $v_0$ along the microtubules, and the induced microtubule 
motion. 
The passive currents, are obtained from the free energy of 
the motor microtubule mixture. It contains 3 terms, the mixing free 
energy of the three components together with an interfacial tension term, the orientational free energy associated to the 
microtubule polarity and a coupling term that describes the effective 
surfactant behavior of the  microtubules at the interface between motor domains, i.e. their tendency to align to motor gradients. Further details of the theoretical model, linear stability analysis, parameter estimates, and numerical simulations are provided in the Supplemental Material~\cite{supplemental}.

The mixing free energy per unit area is given by the classical Flory-Huggins 
theory and reads

\begin{equation}
\begin{gathered}
\frac{f_{mix}}{kT \rho_0}= \hat{c}_+ \log \hat{c}_+ + \hat{c}_-\log \hat{c}_- + \\+\frac{\pi d}{2L}\psi \log \psi + (1- \hat{c}_{tot})\log (1- \hat{c}_{tot}).
\label{eq:f_mix}
\end{gathered}
\end{equation}

It ensures that the sum of the area 
fractions of the three components remains smaller than one, 
$\hat{c}_{tot} < 1$.

In a mean field approximation, the polarization free energy of the microtubules 
per unit area can be written as a Landau expansion in powers of the polarity 
field
\begin{equation}
\begin{gathered}
f_{p} = \frac{k_B T}{2}\phi\left(\beta\phi |\p|^2(1+\frac{\alpha}{2}|\p|^2)+\kappa(\phi 
\nabla.\p)^2\right)
\end{gathered}
\end{equation}
The average density of the microtubules is small enough that the orientation of 
the microtubules remains random 
(except at the interfaces) so that the coefficient $\beta$ is positive. Still we 
keep track of the next term in 
the expansion $\alpha$ to ensure that at the interface between two motor domains 
where the microtubules are strongly aligned, the modulus of the polarity remains 
smaller than one. The final term is the Frank orientational free  energy,
where we have assumed that the Frank bending constants are equal to $\kappa$
and positive \cite{deGennesProst1993}. 

Finally, we add a coupling term. A precise study of its origin requires a detailed microscopic theory that we propose in Section I.7 in SI. The main prediction is that  motors exert a torque that tends to rotate the filament polarization and align it to motor concentration gradients. In our model, the torque derives from an effective energy that, for simplicity,  we include in the free energy of the system. While this torque would be allowed by rotational invariance as long as the system is polar, even if passive, here the polarity of the filaments is only relevant in the presence of walking molecular motors. It follows that this coupling is an active term. The simplest possible form of such active energy coupling the two vectors $\bf p$ and $\nabla c_\pm$ is
\begin{equation}
\begin{gathered}
f_{c} =  - k_B T\nu \phi \p \cdot\grad(\Cp-\Cm)
\label{eq:coupling}
\end{gathered}
\end{equation}
where the amplitude of the torque is measured by an active coefficient $\nu$ and we consider that the two motors are identical except for their 
directionalities along the filaments. The coupling free energy plays a role whenever the concentration gradients are finite, i.e., at the interface between two phases.

The microscopic model in the SI gives an estimate of 
the coefficient
$\nu = \epsilon \frac {2 L d^2 \log{L/d}}{\pi}$, where the dimensionless number $\epsilon$ depends on  $\tilde\epsilon=\frac{k_{on} L}{D}$, where $k_{on}$ is a microscopic binding rate defined in SI, which compares bulk diffusion  and microtubule binding 
for the motors:  $\epsilon= \frac{\pi \tilde \epsilon}{2 \log(L/d)}$ if $\tilde\epsilon< 1$ 
and $\epsilon=1$ if $\tilde \epsilon>1$. Motivated by the experiments on microtubules, the result of the SI has been obtained in a limit where the duty ratio of the motors is high. It gives a value of $\nu$ independent of the motor velocity, but it is not valid in the limit of low velocities. Regardless, $\nu$ is still expected to vanish in the passive limit where  $v_0=0$.

The expression of the total (effective) free energy is then written as 
\footnotesize

\begin{equation}
\centering
\begin{gathered}
    F = \int d\textbf{r} \;[\;f_{mix} 
   +f_p +f_c +\frac{k_B T }{2}\zeta\left(|\grad \Cp|^2+|\grad\Cm|^2\right)\;]
\label{eq:freeenergy}
\end{gathered}
\end{equation}
\normalsize

The chemical potentials of the three components are 
calculated as $\mu_i=\frac{\delta F}{\delta c_i}$, where $c_i$ denotes $\Cp , \Cm, \phi$.

As in active model B of phase 
transitions with conserved order parameters \cite{wittkowski_scalar_2014}, the passive fluxes of the three 
components are proportional to the respective gradients of the chemical potentials: 
${\bf j}^{pass}_i = -c_i A_i \nabla \mu_i$. The transport coefficients are the 
molecular mobilities of the three components, they are related to their diffusion 
coefficients when they are very dilute: $D_{\pm}= k_B T A_{\pm}(=D)$ for the 
motors, where the last equality holds if again the motors are identical except in their directionalities, and $D_{mt}= k_B T A_{mt}$ for the filaments.

The active motor current is due to the directed motion of the motors along the microtubules at 
the velocity $\pm v_0 \bf p$, with the two types of motors moving in opposite directions. However, one must also take into account the fact that the motion of the 
motors creates a motion of the 
microtubules in the opposite direction at a velocity $v_{mt} \bf p$, which it itself advects the motors. 
Hence, the active motor currents are given by $\textbf{j}^{act}_\pm =(\pm v_0+v_{mt})\phi n_\pm\p$
 where $n_\pm$ are the numbers of motors bound to a microtubule of length $L$. The microscopic model presented 
in the SI leads to $n_\pm= \frac{2 D L c_\pm \log(L/d)}{\pi v_0} $.
Thus we write the active motor current\begin{equation}
\begin{gathered}
    \textbf{j}^{act}_\pm = \frac{(\pm v_0+v_{mt})\psi\Cpm}{\Re} \p
\end{gathered}.
\end{equation}
where $\Re={\frac{2\hat{c}_{\pm}L}{\pi n_{\pm}d}}(=\frac{\pi d v_0}{D \log(L/d)})$ is an effective duty ratio of the motors.

\begin{figure}[t]
  \includegraphics[width=\columnwidth]{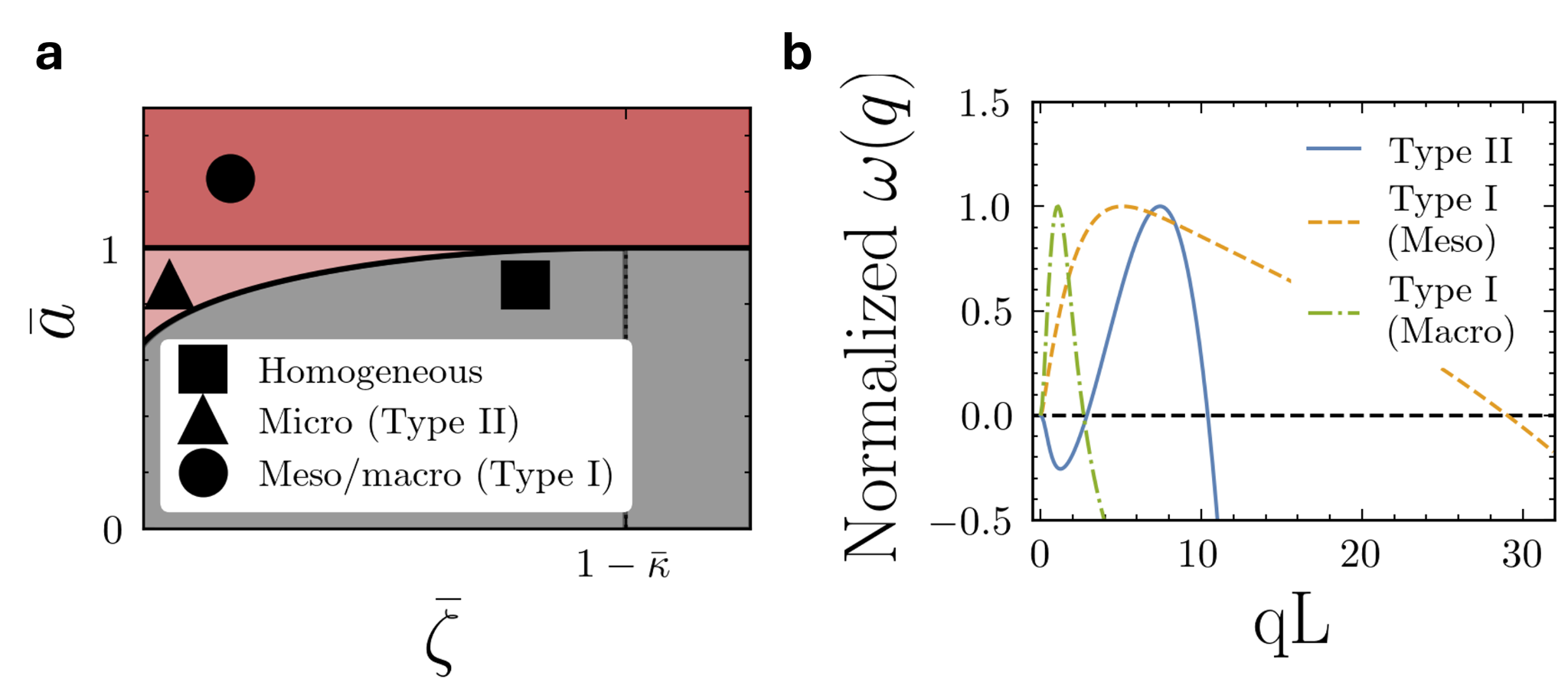}
  \caption{ {\bf a} Stability diagram for  $\bar{\kappa}<1$, in the ($\bar{\zeta}, \;\bar{a}$) plane. Different shaded areas correspond to different phases. We set $\bar{\kappa}=0.56$.
{\bf b} Eigenvalues (eq. S11 in SI) for Type I (dashed lines, green for macro and orange for mesophase separation) and Type II (solid), corresponding to the points marked in (a). The eigenvalues are normalized by their maximum value for clarity. The length scale is rescaled by the microtubule length $L$.}
  \label{fig:stability}
\end{figure}
The active velocity of a microtubule is calculated in Section I.6 in SI from the force balance on the 
microtubule. In the limit where the external friction 
on the microtubules is small: 
$v_{mt}=-v_0\frac{\Cp-\Cm}{\Cp+\Cm}$. The corresponding 
active current of microtubules is: 
\begin{equation}
\begin{gathered}
\textbf{j}_{mt}^{act}= -v_0\frac{\Cp-\Cm}{\Cp+\Cm}\phi\p
\end{gathered}
\end{equation}

The dynamic equations for the concentrations of the 3 components are given by the conservation laws $\frac{\partial c_i}{\partial t} + \nabla {\bf j}_i=0$ where for each component (motors and microtubule), the total current ${\bf j}_i ={\bf 
j}^{pass}_i+{\bf j}^{act}_i$ is the sum of the passive and  active components. 

The microtubule polarity is a non-conserved order parameter. We follow model A of phase 
transitions \cite{bray_theory_2002} and write that the total derivative of the 
polarity is proportional to the orientational field $\bf h =-\frac{\delta F}{\delta \bf p}$. We 
obtain:
\begin{equation}
\label{eq:pola}
\frac{\partial \mathbf{p}}{\partial t} 
+ v_{mt} (\mathbf{p}\cdot \nabla)\mathbf{p} 
= -\frac{1}{\Gamma \phi}\,\frac{\delta F}{\delta \mathbf{p}}   \\
\end{equation}
where $\Gamma$ is the rotational friction coefficient of a microtubule  
that is related to its rotational diffusion constant $D_r=k_B T\beta/\Gamma$ and we have included advection of the polarity with velocity $v_{mt}$.
\footnote{The coefficient $\nu$ measuring the torque created by a concentration gradient on the microtubules appears both in the polarization equation \eqref{eq:pola} and in the expressions of the 3 currents. A more general approach would be to use similar symmetry arguments to the one we use for the effective energy equation directly for the polarization equation and each current. This leads to equations with the same structure as the ones presented here, with however 4 different coefficients replacing $\nu$. We checked by setting $\nu=0$  in the conservation equations for the filaments and the two motors that it leads to qualitatively similar results. For simplicity, we therefore keep here the effective energy approximation where the 4 coefficients are equal to $\nu$.}

This gives a full set of dynamical equations describing each species. We first consider a steady homogeneous and isotropic 
state where the concentrations of motors and 
microtubules are constant and the microtubules are randomly 
oriented ($\p={\bf 0}$). We study the
stability of this state with respect to a perturbation of 
wave vector $\bf q$. We define a five component order 
parameter vector {\bf x}= $(\phi,\Cp, \Cm, {p}_x, { p}_y)$. Its 
value in the steady state is ${\bf x}_0$ and the relaxation  
of a perturbation 
$\delta {\bf x}=\delta {x}_0 \exp i \bf q \cdot \bf r$ follows the equation 
\begin{equation}
    \frac{\partial \delta {\bf x}}{\partial t}= {\bf M} \ \delta {\bf x}
    \label{eq:relaxation}
\end{equation}
The explicit expression of the relaxation matrix is 
given in the SI in the limit where we can neglect the lipid contribution to the free energy. The system is unstable and phase separates if 
the relaxation matrix has positive eigenvalues. For 
a given value of $q=|\bf q|$,
$\textbf{M}$ has only one possibly positive 
unstable eigenvalue $\omega(q)$ and  it can be 
diagonalized numerically to obtain the unstable 
mode.

In the symmetric case $\hat{c}_\pm^0=\hat{c}_0=c_0/\rho_0$, the 
relaxation matrix can be diagonalized explicitly, 
and it allows for a general linear stability analysis. The expression of the  eigenvalue $\omega(q)$ is given in the SI (Eq.~[S11]). 
We use here dimensionless units rescaling lengths 
by the microtubule length $L$ and time by their 
inverse rotational diffusion constant $D_r^{-1}$. 
The behavior of the system depends then on 5 
dimensionless parameters: an effective coupling between 
polarization and concentration gradient $\bar b = 
\frac{\pi D d\rho_0 \nu \Re}{2 v_0 L^3}(=\frac{ d^2}{L^2})$, an activity 
parameter $\bar a=\frac{2 \epsilon^2 \nu v_0 \hat{c}_0\psi_0\rho_0}{\beta \Re D} (=\frac{4 \epsilon^2 \log^2(L/d)\hat{c}_0 \psi_0L}{d \pi^3})$, a diffusion constant $\tilde D =\frac{D}{D_r L^2}$, a parameter related to the 
interfacial tension between the two types of motors 
$\tilde \zeta= \frac{\zeta \rho_0 \hat{c}_0}{L^2}$ and a rescaled 
bending modulus ${\tilde \kappa= \frac{\pi d\rho_0\kappa\psi_0}{2 \beta L^3}}$. The values of $\bar a$ and $\bar b$ inside parenthesis are the values obtained from the microscopic model in the SI.

If one rescales $\bar \zeta = \tilde  \zeta/\bar b$ and $\bar  \kappa=\tilde \kappa/\bar b$, the stability of the system depends only on the values of the three parameters $\bar a$, which measures the activity, and $\tilde \zeta$ and $\tilde \kappa$, which stabilize the homogeneous phase. In the following, we fix  the values of $\bar b$, $\tilde D$ and $\bar \kappa$ and we study 
the stability in a plane $(\bar \zeta, \bar a)$, i.e. varying the effective  interfacial tension and activity.    

There are two possible topologies of the stability diagram. First, if $\bar 
\kappa\geq 1$, the system is unstable at small 
enough wave vectors if $\bar a \geq 1$ and stable 
at all wave vectors if $\bar a \leq 1$. The growth 
rate $\omega (q)$ has a peak corresponding to a 
maximum growing wave vector $q^*$. The microtubule-motor mixture becomes more unstable if the activity ($\bar 
a$) increases and is stabilized at larger wave 
vectors by the bending 
modulus $\bar \kappa$ and by $\bar \zeta$. In the theory of pattern formation \cite{cross_pattern_2009}, this 
case corresponds to a Type I instability where the system is unstable at the macroscopic scale $q \approx 0$.

\begin{figure}[t]
\centering
  \includegraphics[width=\columnwidth]{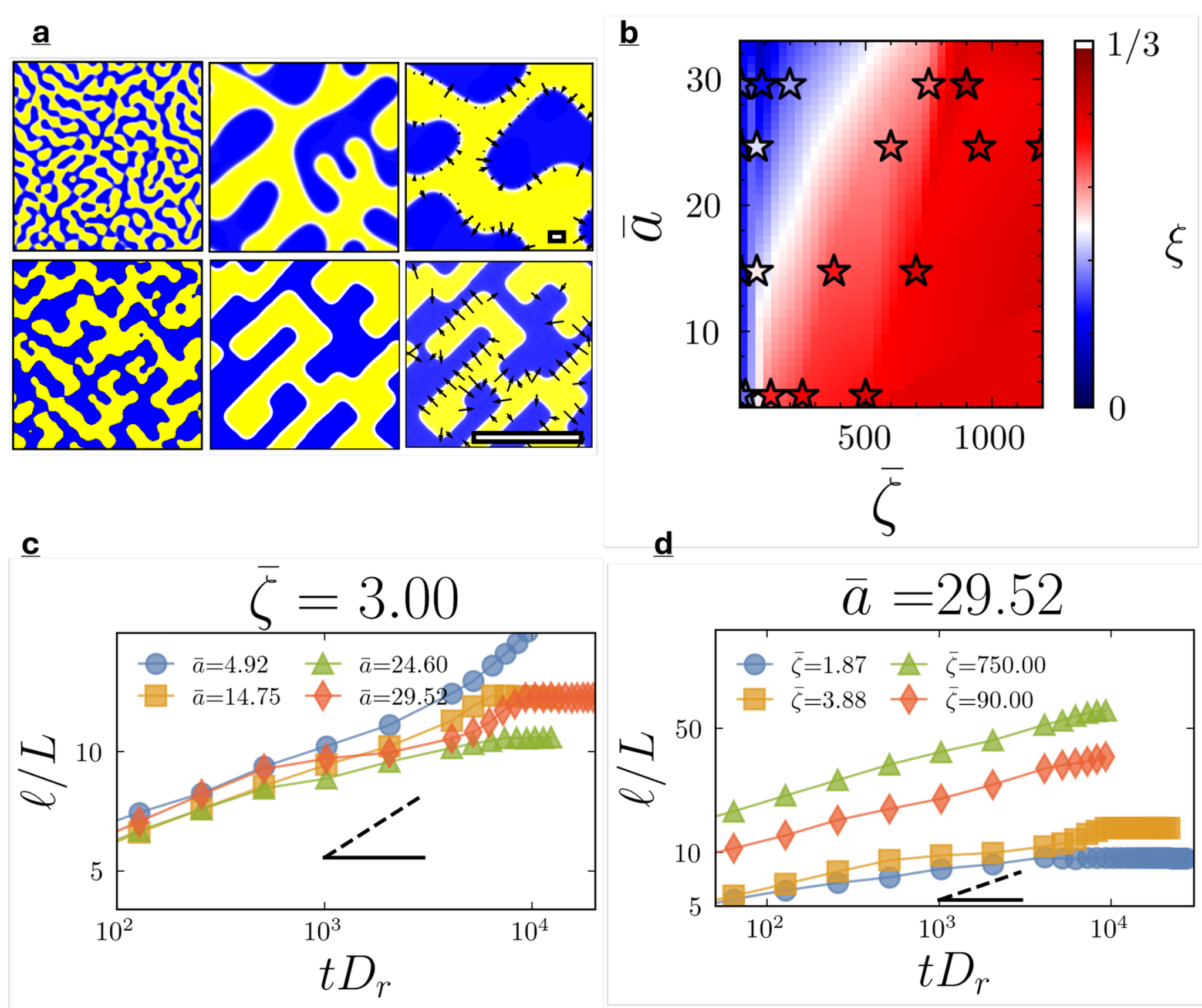}
 \caption{
 {\bf a} Snapshots of simulations ($D_r t=64, 8000, 28000\;$) in the Type I coarsening region (top, ${\bar \zeta}=2800$, $\bar{a}=1.5$), one showing arrested coarsening (bottom, $\bar{\zeta}=1.87$, $\bar{a}=29.52$). Arrows indicate the polarity $\p$, blue is $\Cp$ and yellow is $\Cm$. Scale bars are $10L$, notice the difference in scales. {\bf b} Heatmap of the growth exponent $\xi$ inside the coarsening (Type I) region, showing the slow down of coarsening also where the linear stability analysis would predict full coarsening. The plot is obtained by interpolating simulations, marked by stars {\bf c-d} Length plot $\ell(t)$ at fixed activity ({\bf c}), indicating how high $\bar{\zeta}$ favors coarsening, and at varying activity with fixed $\bar{\zeta}$ ({\bf d})  showing that activity favors arrested coarsening. Black dashed line indicates a $\sim t^{1/3}$ scaling, solid black line is a constant.} 
  \label{fig:sims}
\end{figure}

In the other limit where $\bar \kappa\leq 1$, shown in Fig.~\ref{fig:stability}.a, the system has a 
similar macroscopic instability if $\bar a \geq 1$. But it has another kind of 
instability for $\bar a<1$ and small values of $\bar
\zeta$.  There, the relaxation rate is 
negative for wave vectors smaller than a value $q^{**}$ and has a positive peak at 
a larger wave vector $q^*$  (Fig.~\ref{fig:stability}.b). 
This type II instability occurs if $(\bar \zeta -\bar a+\bar{\kappa})^2 \ge 4 \bar\zeta \bar \kappa(1-\bar a) 
$, and corresponds to an arrested microphase separation. Similar instabilities are found numerically in the case where the two motor concentrations are not equal; examples are given in the SI where we also discuss the fastest growing mode. 

The phase diagram of the system cannot be obtained from the linear stability analysis and requires a numerical 
solution of the full steady state Cahn-Hilliard equations. 
We hence make the equations dimensionless and 
solve them using a finite difference method. The detailed procedure is presented in Section III in SI. Simulations are performed in the $(\bar \zeta, \bar a)$ space. The results are presented in (Fig.~\ref{fig:sims}). We find 3 types of behaviors: a stable homogeneous 
phase where the motors and the microtubules are mixed; macrophase separation where at long times the system phase 
separates into two phases each containing one motor type, with the microtubules forming a polarized 
interfacial layer; and micro/mesophase separation with alternating domains of the two 
motor types separated by an interfacial monolayer of microtubules that grow slowly, i.e. they show almost arrested coarsening (Fig.~\ref{fig:sims}.a, Movies SM1-2).

We estimate the characteristic domain size $\ell(t)$ at  a given time from the structure factor of the motor density (see SI) and we use the growth exponent $\xi$ of $\ell(t)\sim t^{\xi}$ as a proxy for coarsening or arrested separation. The signature of macrophase separation, is an increase of the length scale over time $\ell(t)$ until it reaches the size of the system. As in classical spinodal decomposition, we find that the length $\ell$ increases as a power law  in time with an exponent $\xi\approx 1/3$, which is the exponent predicted for passive systems by the Lifshitz-Slyozov theory \cite{lifshitz_kinetics_1961}. The motors phase separate into two domains. Macrophase separation is obtained for large values of the interfacial parameter $\bar \zeta$ and ${\bar a}>1$.

In the case of micro/mesoscopic phase separation, coarsening stops and the length scale $\ell(t)$ saturates at long times to a finite value ($\xi\approx 0$), which is the size of the domains. For symmetric motors at the same concentration, in most cases we observe a disordered bicontinuous mesophase of the two motor domains. Microphases are formed in the parameter domain where the linear stability analysis  predicts a type II instability with a characteristic length slightly larger than the one corresponding to the peak position at the instability threshold. In this case the characteristic length  is selected by the non-linearities.  However the domain of existence of mesophases is even larger, and we also find arrested coarsening for ${\bar a} > 1$ and $\bar{\zeta}$ small (Fig.~\ref{fig:sims}.b).  In short, high activity pushes the system towards mesophase separation, provided that the interfacial term is not too high. This is further shown in Fig.~\ref{fig:sims}.c-d where the behavior of $\ell(t)$ is shown in different conditions (with $\bar{a}>1$) varying independently $\bar \zeta$ or $\bar a$ to make the system transition from meso- to macrophase separation. 
  
We then varied the relative motor concentration and we found droplet phases of one motor type in a continuous phase of the other motor, stabilized by a microtubule interface ( Fig.S1(a) in SI). When the concentration ratio between the two types of motors is too large, the system does not phase separate but there is a macroscopic c active current of microtubules  This corresponds experimentally to the state in which microtubules just glide propelled by the dominant motor\cite{utzschneider_force_2024}. 

In conclusion, this paper introduces a new type of non-equilibrium phase separation where two components (the two motors) are actively and differentially sorted out in space by a third component (the microtubules), and reports how this results in domains enriched in either component that, in some conditions, do not coarsen. Phase separation by transport or by active pumps is a new observation, and one that could play a role in the organization, e.g., of the cell's cytoplasm and of its cytoskeleton.

We only present here a phase diagram where the microtubule concentration is fixed, i.e., a projection of the phase diagram in the plane spanned by the concentrations of the two motors.  Our active Cahn-Hilliard theory gives a precise theoretical description of this phase separation and  shows the richness of the phase diagram that includes gliding/homogeneous phases, macroscopic and mesoscopic phase separation.  However, due to the finite length of the simulation, it is not always easy to distinguish between a steady state mesophase and a system slowly coarsening toward a macroscopic phase separation and showing a (non-equilibrium) spinodal decomposition pattern. 
It is also important to note that the phase separation is driven by the polar transport of motors by the microtubules. If the motor velocity vanishes, the 3 components do not phase separate. This equilibrium limit does not appear clearly in our equations because we have assumed that the effective duty ratio and therefore the velocity are large enough in order to remain closer to experiments.

Future experiments could provide stronger tests of the theory by studying the change in the phase diagram with the variation of the microtubule length and/or of the effect of relative motor concentration and speed. From a more theoretical point of view, our study remains at the mean field level  and both thermal and active fluctuations could play a role. One could also study the existence of multi-critical points for the mixed, macroscopic and microscopic phases.
\\

\section{Data availability}
All codes used for this manuscript is available at https://github.com/Sudiphys/Phase-separation-by-polar-active-transport.

\section{Author contributions}
SP and AS designed research, developed the theory, analysed results. SP performed simulations. JFJ designed research and developed the theory. AS and JFJ wrote the text, SP wrote the Supplementary Information.  MT and LB designed research. All authors revised the manuscript.

\section{Acknowledgments}
AS has received funding from the European Union’s Horizon 2020
research and innovation programme under the Marie Skłodowska-Curie grant
agreement No 101108326. AS thanks Filippo De Luca and the group of Erwin Frey for helpful discussion. 

\nocite{hohenberg_theory_1977,Jiang2019}
\bibliographystyle{apsrev4-1} 
\bibliography{biblio} 

\end{document}


\title{Supplemental Material for: Phase Separation by Polar Active Transport}

\author{Sudipta Pattanayak}
\email{These authors contributed equally}
\affiliation{Collège de France, Université Paris Sciences et Lettres, Matière molle et biophysique, Paris 75231, France
}
\affiliation{Institut Curie, Université Paris Sciences et Lettres, Physique de la Cellule et Cancer, Paris Cedex 05 74248, France}
\author{Alfredo Sciortino}
\email{These authors contributed equally}
\affiliation{CytoMorpho Lab, Chimie Biologie Innovation, UMR8132, Université Paris Sciences et Lettres, Ecole Supérieure de Physique et Chimie Industrielles de la Ville de Paris, CEA, CNRS, Institut Pierre Gilles De Gennes, Paris 75005, France}
\author{Laurent Blanchoin}
\affiliation{CytoMorpho Lab, Laboratoire de Physiologie Cellulaire et Végétale, UMR5168, Université Grenoble-Alpes, CEA, INRA, CNRS, Interdisciplinary Research Institute of Grenoble, Grenoble 38054, France}
\affiliation{CytoMorpho Lab, Chimie Biologie Innovation, UMR8132, Université Paris Sciences et Lettres, Ecole Supérieure de Physique et Chimie Industrielles de la Ville de Paris, CEA, CNRS, Institut Pierre Gilles De Gennes, Paris 75005, France}
\author{Manuel Théry}
\affiliation{CytoMorpho Lab, Laboratoire de Physiologie Cellulaire et Végétale, UMR5168, Université Grenoble-Alpes, CEA, INRA, CNRS, Interdisciplinary Research Institute of Grenoble, Grenoble 38054, France}
\affiliation{CytoMorpho Lab, Chimie Biologie Innovation, UMR8132, Université Paris Sciences et Lettres, Ecole Supérieure de Physique et Chimie Industrielles de la Ville de Paris, CEA, CNRS, Institut Pierre Gilles De Gennes, Paris 75005, France}
\author{Jean-Fran\c cois Joanny}
\email{Corresponding, email: jean-francois.joanny@college-de-france.fr}
\affiliation{Collège de France, Université Paris Sciences et Lettres, Matière molle et biophysique, Paris 75231, France
}
\affiliation{Institut Curie, Université Paris Sciences et Lettres, Physique de la Cellule et Cancer, Paris Cedex 05 74248, France}

\maketitle

\renewcommand{\theequation}{S\arabic{equation}}
\setcounter{equation}{0}

\renewcommand{\thefigure}{S\arabic{figure}}
\setcounter{figure}{0}

In the Supplementary information, we explain the details of the active Cahn-Hilliard theory and the numerical procedure used to study phase separation in a three-component system in two-dimensional space, where we observe the spatial sorting of two non-interacting components by a third polar component. In section \ref{Active_CH_Theory}, we discuss the theoretical model and perform linear stability analysis. Then, we estimate  the parameters from experiments in section \ref{parameter_estimation} and give details of the numerical study in section \ref{numerical_details}, respectively.

\section{Active Cahn-Hilliard theory \label{Active_CH_Theory}}
In this section, we build up an active Cahn-Hilliard theory for the 
experiment described in the main text. We first discuss the effective free energy  and the passive and active currents of motors and microtubules that 
drive the dynamics of the mixture.

\subsection{Free energy}
The model considers three species, the "microtubules", with polarity $\p$ and two "motor types", that walk along microtubules each in one direction $\pm$ and induce motion of the microtubules.
The motors and the microtubules are embedded in a fluid membrane where they can diffuse.

The effective thermodynamics requires 3 conserved fields: the surface 
concentrations of the two motor types  $\cpm$, where the motors labeled 
$+$ and $-$ motors walk along the microtubules in the $\pm \p$ directions  
respectively, and the microtubule surface concentration $\phi$. The microtubules' polarity $\p$ is defined as their local average orientation. The polarity $\p$ is a non-conserved order parameter. 
The two types of motors 
walk along the microtubules' polarity with respective velocities ${\bf 
v}_{\pm}=v_{\pm} \p$

We assume that motors of the two types are identical in all respects except for the direction they move along the microtubules. This assumption can be easily relaxed allowing for different motor velocities or diffusion constants but we keep it for simplicity, so that $v_+=-v_-   =v_0$.

The thermodynamics of the motor-microtubule mixture is described by a  free energy $F$ per unit area. We use here a Flory-Huggins theory for an incompressible quaternary mixture where the 4th component is the fluid lipid membrane that we consider as a solvent. We ignore any interaction terms between motors and microtubules (all the Flory interaction parameters vanish) and keep only the mixing entropy contribution.  We include terms proportional to the square of the gradients of the motor concentrations that act as an interfacial energy term. We suppose that the 
total system including the membrane is incompressible. Note that this is 
equivalent to introducing excluded volume between the components. We 
introduce the surface fractions of the 3 components $\hat{c}_\pm =\pi d^2 c_\pm =c_\pm/\rho_0$ 
and  $\psi= 2Ld\phi$ where $d$ is the motors' radius, $L$ is the microtubules' length and $\rho_0 =1/\pi d^2$ is a reference concentration. Microtubules have a thickness $2d \ll L$. We hence require that $\psi+\hat{c}_++\hat{c}_-<1$. 
We also introduce a term that tends to align the microtubules' polarity to the gradients of motor concentrations, as if they were surfactants. The structure of this term is constructed using dimensional arguments given in the main text and, while it is included here in the free energy, it has a non equilibrium origin. 
Finally the free energy contains an orientational contribution of the polarity that would lead to a non-polar phase in the absence of motor gradients.
The free energy reads

\begin{equation}
\\\frac {F}{k_B T} =\int d\textbf{r}\left(\frac{f_{mix}}{k_B T}(\cpm, \phi)+\frac{\beta}{2}\phi 
|\p|^2(1+\frac{\alpha}{2}|\p|^2)+\frac{\kappa}{2}(\phi \nabla.\p)^2+\frac{\zeta}{2}(|\grad \Cp|^2+|\grad\cm|^2)- \nu \phi \p \cdot\grad(\Cp-\cm)\right)
\end{equation}

The entropic contribution $f_{mix}$ is given by:
\begin{equation}
\frac{f_{mix}}{kT \rho_0}= \hcp \log \hcp + \hcm\log \hcm + \frac{\pi d}{2L}\psi \log \psi + (1- \hcp -\hcm - \psi)\log (1- \hcp -\hcm - \psi)
\end{equation}

The polarity contribution to the  free energy $\left(\frac{\beta}{2}\phi 
|\p|^2(1+\frac{\alpha}{2}|\p|^2)\right)$ is obtained from a mean 
field Landau-Ginzburg expansion of the free energy. The two coefficients $\alpha$ 
and $\beta$ are positive and favor a random orientation of the 
microtubules. The Frank term $\left(\frac{\kappa}{2}(\phi \nabla.\p)^2\right)$, is written for simplicity in the one-constant approximation, with a positive bending constant $\kappa$. The bending energy is proportional to $\phi^2$ because it is due to interactions between different microtubules.

The term proportional to the square of motor gradients $\left( \frac{\zeta}{2}(|\grad \Cp|^2+|\grad\cm|^2)\right)$ is at the origin of the interfacial tension between the two motors when they phase separate.

Finally, the last term of the free energy is the mechanical energy associated to a 
torque that aligns the polarity  to the concentration gradient. 
The torque is calculated within a microscopic model in section \ref{sec:nmotors}. The torques induced by the two types of motors 
have opposite signs. As discussed in the main text, this is an active contribution to the energy that we 
include in the effective free energy for convenience. 
 
\subsection{Passive and active currents}
\subsubsection{Passive currents}

In the following, we denote by $k_B T \chi_{ij}$ the matrix of the second derivatives of the free energy (the osmotic compressibility matrix).

Following the lines of model B of phase transitions \cite{hohenberg_theory_1977} with a conserved order parameter, we use the Onsager approach where the current of each species is proportional to the gradient of its chemical potential, ignoring therefore  non-diagonal couplings. In the simplest model, the mobility is proportional to the local concentration. We therefore write the currents as 
\begin{equation}
    {\bf j}_i=- A_i c_i\nabla \mu_i  = - A_i c_i \nabla \dF{c_i}
\end{equation}
where the index $i$ refers to the three components (the two motors and the microtubules) and $c_i$ to their concentrations. The positive quantities $A_i$ are the molecular mobilities. 

Polarity is a non-conserved order parameter and as in model A \cite{hohenberg_theory_1977} of phase transitions, the Onsager prescription is that the time derivative of the order parameter is proportional to its conjugate field
\begin{equation}
    \frac{\partial {\bf p}}{\partial t}=-\frac{1}{\Gamma\phi} \frac{\delta F}{\delta {\bf p}}
    \label{eq:polarity}
\end{equation}
where $\Gamma$ is the rotational friction coefficient per molecule. Note that as we have included the active torques in the effective free energy, this relation takes into account both passive and active contributions.

\subsubsection{Active currents}
The motion of the motors on the microtubules induces an active current of the microtubules at a velocity $v_{mt}$ along their local polarity. The motors themselves are convected by the microtubules. The active current of motors is
$\textbf{j}_{act}^\pm =(\pm v_0+v_{mt})\phi n_\pm \p$
where $n_\pm$ are the numbers of motors bound to a microtubule given by Eq. ~\eqref{eq:number}, so that the active motor current is 
\begin{equation}
\begin{gathered}
\textbf{j}^{act}_\pm = \frac{({\pm}v_0+v_{mt})\psi\cpm}{\Re} \p
\end{gathered}
\end{equation} 
where $\Re$ is defined in the main text.

The microtubule current is given by  
\begin{equation}
\textbf{j}^{act}_{mt}=v_{mt}(\p, \cpm)\phi\p=\rho_0\pi(d/(2L))v_{mt}(\Cp, \cm)\psi\p.
\end{equation} 
The velocity of a microtubule is obtained from the force balance on a system formed by the microtubule and its bound motors in section \ref{sec:mtvelocity}.  We take here the limit where the external friction on the microtubule is small:
\begin{equation}
v_{mt}=-v_0\;\frac{c_+-c_-}{c_++c_-}
\end{equation}

\subsection{Dynamic equations}
The dynamic equations are obtained by writing the conservation laws for each component $\frac{\partial c_i}{\partial t} + \nabla \cdot{\bf j^t}_i=0$ where ${\bf j^t}_i$ is the total current (active plus passive),
and the dynamic equation of the polarity is derived from Eq:~\eqref{eq:polarity}

The explicit form of the dynamic equations is:
\begin{subequations}
\begin{eqnarray}
\frac{d{\cpm}}{dt} &=\Div \left((\mp v_0-v_{mt}(\Cp, \cm))\psi\cpm\p\right)+c_+ \chi_{\pm\pm} D\Delta\cpm +\cpm \chi_{\pm\mp} D\Delta c_{\mp}+\cpm \chi_{\pm\phi} D\Delta \phi \\
&+D\Div \left(-\zeta\cpm\grad\Delta\cpm\pm\nu  \cpm\grad\Div (\phi\p)\right)
\end{eqnarray}
\begin{eqnarray}
\frac{d\phi}{dt} &=\Div (-v_{mt}(\Cp, \cm)\phi\p)+D_{mt}\left(\phi \chi_{\phi\phi} \Delta \phi +\phi \chi_{\phi +} \Delta c_+ + \phi \chi_{\phi -} \Delta c_-\right)\\
&+D_{mt}\Div \left(\frac{\beta}{2}\phi \grad(|\p|^2(1+\frac{\alpha}{2} |\p|^2))+\kappa\phi\grad(\phi(\Div\p)^2)-\nu \phi\grad(\p\cdot \grad(\Cp-\cm))\right)\\
\end{eqnarray}
\begin{equation}
\frac{d\p}{dt}+ v_{mt} (\mathbf{p}\cdot \nabla)\mathbf{p}  = -\frac{1}{\Gamma} \left( \beta \p (1+\alpha |\p|^2)-\frac{\kappa}2 \grad (\phi^2 \Div\p)-\nu \grad (\Cp-\cm) \right)
\end{equation}
\end{subequations}
where we have introduced the diffusion coefficients of the motors $D=A k_B T$ (having set $A_+=A_-=A$ since the motors are considered as identical) and of the microtubules $D_{mt} = A_{mt} k_B T$ and the inverse compressibility matrix of the free energy $\chi$ (the matrix of second derivatives). The rotational diffusion constant of the polarity is 
$D_r=\dfrac{k_B T\beta}{\Gamma}$. 

Note that the diffusion terms in the equations for the concentrations are rather complicated, including cross-terms and effective diffusion constants that contain thermodynamic factors. In the limit of small concentrations of the motors and microtubules, the cross diffusion constants are negligible and the effective diffusion constants are equal to the bare diffusion constants since $c_i \chi_{ii}=1$. In the following, we use this simple limit for the stability analysis but we keep all the terms to study numerically the phase separation, because the area fraction $\psi$ of the microtubules can become close to $1$ at an interface between the two motor types. 

The equations are made dimensionless by introducing the area fraction of the two components $\cpm=\hcpm \rho_0$ and $\phi=\psi \rho_0\pi d/(2L)$ where $\rho_0=1/(\pi d^2)$ is the inverse of the area of a motor and $L$ is the length of a microtubule. We also introduce the reduced length and timescale $r^*=r/L$ and $t^*=D_rt$.

\begin{subequations}
\begin{equation}
\begin{gathered}
\frac{d{\hcpm}}{dt^*} ={\frac 1 {D_r L}}\aDiv\left((\mp v_0-v_{mt}) \frac{1}{\Re}\psi\hcpm\p\right)+{\frac D{D_r L^2}} \Delta^*\hcpm+\frac{ D}{D_r L^2} \aDiv\frac{\hcp}{1-\hcp-\hcm-\psi}\grad^*\hcp
\\ +{\frac D{D_r L^2}} \aDiv\left(-{\tilde \zeta}\frac{\hcpm} {\hat{c^{\pm}}_0}  \grad^*\Delta^*\hcpm\pm   {\frac{\nu \rho_0 \pi d}{{2}L^2}}\hcpm\grad^*\aDiv(\psi\p)\right)
\end{gathered}
\end{equation}
\begin{equation}
\begin{gathered}
\frac{d\psi}{dt^*} =\frac 1{D_r L}\aDiv (-v_{mt}\psi\p)+\frac{D_{mt}}{D_r L^2} \Delta^* \psi +\frac{ D_{mt}}{D_r L^2} (\frac{2L}{\pi d})\aDiv\frac{\psi}{1-\hcp-\hcm-\psi}\grad^*\psi +{\frac{D_{mt}}{D_rL^2}}\aDiv \beta\Big(\frac{1}{2}\psi \grad^*(|\p|^2(1+\frac{\alpha}{2} |\p|^2)) \\ +\frac{\rho_0 \pi d \kappa}{2L^3 \beta}\psi\grad^*(\psi(\aDiv\p)^2) -\frac{\nu  \rho_0\psi }{ L\beta }\grad^*(\p\cdot \grad^*(\hcp-\hcm)) \Big)
\end{gathered}
\end{equation}
\begin{equation}
\frac{d\p}{dt^*} + {\tilde{v}_{mt}(\p\cdot\nabla^*)\p} = -\left(\p (1+\alpha |\p|^2)-\frac{\rho_0 \pi d \kappa}{2\beta L^3} {\left(2(\Div^*\p)(\grad^*{\psi})+{\psi}\grad^*\aDiv\p\right)}-\frac{\nu\rho_0 }{\beta L} \grad^* (\hcp-\hcm) \right)
\end{equation}
\label{adim-equations}
\end{subequations}

These equations can be rewritten introducing adimensional parameters as
\begin{subequations}
\begin{equation}
\begin{gathered}
\frac{d{\hcpm}}{dt^*} =\aDiv\left(\frac{(\mp {\tilde v_0}-{\tilde v_{mt}})}{\Re}\psi\hcpm\p\right)+{\tilde D} \Delta^*\hcpm+{\tilde D} \aDiv\frac{\hcpm}{1-\hcp-\hcm-\psi}\grad^*\hcpm \\ +{\tilde D} \aDiv\left(-2{\tilde \zeta}\frac{\hcpm}{\hat{c}^+_0+\hat{c}^-_0}\grad^*\Delta^*\hcpm\pm \frac{{\tilde \nu}}{{\tilde v_0}} \frac{\pi d}{2L}\hcpm\grad^*\aDiv(\psi\p)\right)
\end{gathered}
\end{equation}
\begin{equation}
\begin{gathered}
\frac{d\psi}{dt^*} =\aDiv (-{{\tilde v}_{mt}}\psi\p)+\tilde{D}_{mt} \Delta^* \psi+ {\tilde D}_{mt}\aDiv \left( \frac{2L}{\pi d} \right)\frac{\psi}{(1-\hcp-\hcm-\psi)}\grad^*\psi \\
+\tilde{D}_{mt}\beta \aDiv \Big(\frac {1}{2}\psi \grad^*(|\p|^2(1+\frac{\alpha}{2} |\p|^2))+{\tilde \kappa}\frac{\psi}{\psi_0}\grad^*(\psi(\aDiv\p)^2)-\frac{\tilde \nu}{\beta {\tilde v_0}}{\psi}\grad^*(\p\cdot \grad^*(\hcp-\hcm)) \Big)
\end{gathered}
\end{equation}
\begin{equation}
\frac{d\p}{dt^*}  = -{{\tilde v}_{mt}(\p\cdot\nabla^*)\p}-\p (1+\alpha |\p|^2)+{\tilde \kappa}\left(2(\Div^*\p)(\grad^*\frac{\psi}{\psi_0})+\frac{\psi}{\psi_0}\grad^*\aDiv\p\right)+\frac{\tilde \nu}{\beta {\tilde v_0}} \grad^* (\hcp-\hcm) 
\end{equation}
\label{adim-equations}
\end{subequations},

where $c^\pm_0$ is the mean motor concentration, $\psi_0$ is the mean microtubule area fraction and we introduced the parameters

$$\hat{c}_0=\frac{(\hat{c}^+_0+\hat{c}^-_0)}{2}$$
$${\tilde v_0}=v_0/D_rL\;\;\;\;{\tilde v_{mt}}=v_{mt}/D_rL$$
$${\tilde \kappa}=\frac{\pi d \rho_0 \kappa \psi_0}{2\beta L^3}$$
$${\tilde \nu}=\frac{\nu \rho_0{\tilde v_0}}{L}$$
$${\tilde \zeta}=\frac{\zeta \rho_0 \hat{c}_0}{L^2}$$
$${\tilde D}=\frac{D}{D_r L^2}\;\;\;\;\tilde D_{mt}=\frac{D_{mt}}{D_r L^2}$$

\subsection{Linear stability analysis}
We perform a linear stability analysis of the  dynamic equations.
We linearize the equations around a homogeneous state $\hat{c}_\pm=\hat{c}_\pm^0+\delta \hat{c}_\pm^0$, $\psi=\psi_0+\delta\psi_0$ and $\p=\textbf{0}+\delta\p$ and rewrite the linearized equations in matrix form as 

\begin{equation}
\frac{\partial \delta{\bf x}}{\partial t} = \textbf{M}\;\delta{\bf x}
\end{equation}

We combine here the fluctuations into a vector $\delta{\bf x}(\mathbf{r^*},t^*)$ with components
$\{ \delta\hat{c}_+^0, \delta\hat{c}_-^0 ,\delta\psi, \delta p_x, \delta p_y\}$ and consider a periodic perturbation of the dimensionless wave vector $\bf \tilde{q}$: $\delta {\bf x}(\mathbf{r^*},t^*)= \delta {\bf x_0}\exp(-i \mathbf{\tilde{q}}.\mathbf{r^*})$.
We obtain a dimensionless stability matrix ${\bar {\bf M}}$: 
\footnotesize

$$\begin{gathered} \bar{\textbf{M}}=
\begin{pmatrix}
-\tilde {D}\tilde{q}^2(1+\tilde {\zeta} \tilde{q}^2) & 0  & 0 & -i{\tilde q}_x \hat{c}_+^0\psi_0[\frac{\pi d}{2L{\tilde v_0}}\tilde {D}\tilde {\nu} (\tilde{q})^2 & -i\tilde{q}_y \hat{c}_+^0\psi_0[\frac{\pi d}{2L{\tilde v_0}}\tilde {D} \tilde {\nu} (\tilde{q})^2\\
&&&+(\tilde {v}_0+\tilde {v}_m)/\Re]&+(\tilde {v}_0+\tilde {v}_m)/\Re]\\
&&&&\\
0 & -\tilde {D}\tilde{q}^2(1+\tilde {\zeta}  \tilde{q}^2) & 0 & i{\tilde q}_x \hat{c}_-^0\psi_0[\frac{\pi d}{2L{\tilde v_0}}\tilde {D}\tilde {\nu} \tilde{q}^2+ & i\tilde{q}_y \hat{c}_-^0\psi_0[\frac{\pi d}{2L{\tilde v_0}}\tilde {D} \tilde {\nu} \tilde{q}^2+\\ 
&&&+(\tilde {v}_0-\tilde {v}_m )/\Re]&+(\tilde {v}_0-\tilde {v}_m)/\Re]\\
&&&&\\
0 & 0 & -\tilde{q}^2 \tilde {D}_{mt}& -i\tilde{q}_x\psi_0\; \tilde {v}_{mt}(\hat{c}_+^0, \hat{c}_-^0) &-i\tilde{q}_y\psi_0\;\tilde {v}_{mt}(\hat{c}_+^0, \hat{c}_-^0) \\  
&&&&\\
&&&&\\
i{\tilde q}_x \tilde {\nu}/(\beta {\tilde v_0})  & -i\tilde{q}_x \tilde {\nu}/(\beta {\tilde v_0}) &  0  & -1-(\tilde {\kappa}/\beta)  \tilde{q}^2_x & -(\tilde {\kappa}/\beta)\tilde{q}_x\tilde{q}_y \\
&&&&\\
&&&&\\
i\tilde{q}_y \tilde {\nu}/(\beta {\tilde v_0}) &  -i\tilde{q}_y \tilde {\nu}/(\beta {\tilde v_0}) &  0  & -(\tilde {\kappa}/\beta)\tilde{q}_x\tilde{q}_y &-1-(\tilde {\kappa}/\beta) \tilde{q}^2_y \\ \end{pmatrix} 
\end{gathered},$$
\normalsize

where $\tilde{q}=qL$, the adimensionalized wave vector.

Under the assumption that the motors have equal concentrations, $\hat{c}_+^0=\hat{c}_-^0=\hat{c}_0$, we look for the eigenvalues of the matrix $\tilde  M$. We find that only one eigenvalue can become positive and lead to an instability,

\begin{equation}
\omega(\tilde{q})=-\frac{1+[\tilde {D} \tilde{q}^2(1 +\tilde {\zeta}\tilde{q}^2)  + \tilde {\kappa} \tilde{q}^2]}2+ \frac{\left(\Big(1+[\tilde {D} \tilde{q}^2(1 +\tilde {\zeta}\tilde{q}^2)  + \tilde {\kappa} \tilde{q}^2]\Big)^2 
    -4\tilde{q}^2(\tilde{D} + \tilde{D}\tilde{\zeta}\tilde{q}^2 +  \tilde{D}\tilde{\kappa}\tilde{q}^2+ \tilde{D}\tilde{\kappa}\tilde{\zeta}\tilde{q}^4)+4\bar{a} \tilde {D} \tilde{q}^2 [1+\bar{b} \tilde{q}^2]\right)^{1/2}}{2}
 \label{eigenvalue2}
\end{equation}

with the dimensionless numbers $\bar b$ and $\bar a$ defined as

\begin{equation}
\bar{b}=\frac{\pi d}{2L}\frac{{\tilde D}{\tilde \nu}\Re}{\tilde v_0^2}=\frac{\pi D d \rho_0 \nu \Re}{2 v_0 L^3} = \frac{d^2}{L^2}
\end{equation}

and
\begin{equation}
\bar a=\psi_0\hat{c_0}   \frac{\pi d}{L\beta}\frac{{\tilde \nu}^2}{{\tilde v_0}^2\bar b}=\frac{2  \nu v_0 \hat{c}_0\psi_0\rho_0}{\beta \Re D}.
\end{equation}

The system is unstable if the relaxation rate $\omega(\bar{q})$ is positive and stable if it is negative. This leads to the general condition for instability
\begin{equation}
    {\tilde\kappa}{\tilde \zeta} \tilde{q}^4 +({\tilde \zeta} + {\tilde \kappa}- {\bar a}{\bar b})\tilde{q}^2 +(1-{\bar a})=0
    \label{eq:stab}
\end{equation}
In the case where the activity parameter $\bar a$ is smaller than one the system is stable at low wave vector and the instability occurs at finite wave vector. The system has a type II instability. Eq.~\eqref{eq:stab} is a quadratic in $q^2$ and at the instability threshold the two roots are equal. This leads to the stability condition 
\begin{equation}
   ({\tilde \zeta} + {\tilde \kappa}- {\bar a}{\bar b})^2 =4{\tilde\kappa}{\tilde \zeta}(1-{\bar a})
\end{equation}
The wave vector at the instability threshold is then given by 
\begin{equation}
    \tilde{q}^{**2}=\frac{{-\tilde \zeta} - {\tilde \kappa}+ {\bar a}{\bar b}}{2{\tilde\kappa}{\tilde \zeta}}
\end{equation}

In the case where the activity parameter $\bar a$ is larger than one the relaxation rate is positive at small wave vectors, and has a maximum at a finite wave vector and becomes negative for large wave vectors. The system has a type I instability. Close to $\bar a=1$, the most unstable wave vector can be calculated by expanding the relaxation rate at low wave vector
\begin{equation}
  \tilde{q}^{*2}=\frac{\bar a -1}{\tilde \zeta +\bar b [ \tilde \kappa -\bar b +(\bar{a}-1)\tilde D]} 
\end{equation}

\subsection {Number of motors bound to a microtubule}\label{sec:nmotors}

In this section, we consider a microtubule fixed at a given position by an external force for a system containing only one type of motors, say $+$ end directed, and we calculate the number of motors $n$  bound on the microtubule. The microtubule is a thin filament of length $L$. It lies along the horizontal axis (x-axis) from $(x,y)=(0,0)$ to $(x,y)=(L,0)$, where $y$ is the  perpendicular coordinate. The concentration of bound motors per unit length along the microtubule is $\rho(x)$, and the bound motors move along the filament at a constant speed $v_0$. The average concentration of free motors  far from the microtubule is $c_0$. The free motors bind with the filament with a rate $k_{on}$ where $k_{on}$ has dimension of velocity in two dimensions. The  concentration of unbound motors follows the diffusion equation:
\begin{equation}
    D\Big(\frac{\partial^2 c(x,y)}{\partial x^2}+\frac{\partial^2 c(x,y)}{\partial y^2}\Big)=-\rho(L)v_0\delta(y)\delta(x-L)+2 k_{on}c_m(x)\delta(y)H(x),
\label{eq:torque1}
\end{equation}
where $\delta$ denotes the Dirac delta function, H(x) is the characteristic function of the segment $[0,L]$ (equal to $1$ if $x$ belongs to the segment $[0,L]$ and $0$ outside this segment), and $c_m(x)=c(x,y=0)$ is the concentration of unbound motors on the line $y=0$ of the filament. The first term on the right hand side is the flux of motors at the positive end of the microtubule into the external medium and the second term is the flux of adsorbing motors on the microtubules at position $x$. For the sake of simplicity, we ignore here the direct unbinding rate of the microtubule $k_{off}$, which is legitimate if $k_{off}\ll v_0/L$
The integration of Eq. \ref{eq:torque1} between $y=0^-$ and $y=0^+$ leads to the boundary condition 
\begin{equation}
    2D \frac{\partial c}{\partial y}|_{y=0^+}=-\rho(L) v_0 \delta(x-L)+ 2 k_{on} H(x)c_m(x)
    \label{eq:torque2}
\end{equation}
The conservation equation of the bound motors in a steady state reads
\begin{equation}
    v_0 \frac{\partial \rho}{\partial x}=2k_{on}c_{m}(x)
    \label{eq:boundm}
\end{equation}
Taking into account the fact that the concentration of bound motors $\rho(0)=0$  (because there is no flow at the minus end of the filament) and integrating the equation over x, we obtain $\rho(L)= \frac{2 k_{on}}{v_0}\int dx H(x) c_m(x)$.
In order to solve these equations, we introduce one-dimensional Fourier transforms, defining the Fourier transform of a function $f(x)$ as $\bar{f}(q)=\mathcal{F}[f]
(q)=\int_{\mathbb{R}} dx  e^{-iqx}f(x)$.
We define $\delta c(x,y)=c (x,y)-c_0$ and  take the Fourier transform of Eq.~\ref{eq:torque1}. The solution of Eq.~\ref{eq:torque1}  is 
$\delta \bar{c}(q,y)=A e^{-{|q|}y}+B e^{+{|q|}y}$, where $A$ and $B$ are two integration constants. 
Imposing that $\bar{c}(q,y)$ vanishes as $y \to \infty$, we obtain
\begin{equation}
\delta \bar{c}(q,y)=\delta \bar{c}(q,y=0)e ^{-y |q| sgn(y)}
\label{eq:s23}
\end{equation}
The boundary condition Eq.~\eqref{eq:boundm} is written in Fourier space as:
\begin{equation}
    2 D \frac{d \delta \bar{c}}{d y}|_{0+}=-\rho(L)v_0 e^{-iqL} + 2 k_{on}\mathcal{F}[c_m(x)H(x)]
    \label{conccentration2}
   \end{equation}
So that the Fourier transform of the concentration of motors is  
\begin{equation}
    \delta \bar{c}(q, y=0)=\frac{1}{2|q|D}\Big( v_0 \rho(L)e^{-iqL}-2 k_{on} \mathcal{F}[H(x)c_m(x)]\Big)
    \label{torque5}
\end{equation}
We obtain the concentration profile at $y=0$ by inverting the Fourier transform
\begin{equation}
 \delta c(x, y=0)=\frac{1}{2D} \int_{\mathbb{R}} dx^\prime K(x-x^\prime)\Big(v_0 \rho(L) \delta(x^\prime-L)-2 k_{on} H(x^\prime) c_m(x^\prime) \Big)
 \label{torque3}
\end{equation}
where we have defined the convolution kernel $K(x)=\mathcal{F}^{-1}[\frac{1}{|q|}](x)=-\frac{1}{\pi}ln\frac{|x|}{C}$, where $C$ is an integration constant. 
Using the expression of the density $\rho (L)$, we rewrite this equation as 
\begin{equation}
    \delta c(x,y=0)=-\frac{k_{on}}{\pi D} \int_{\mathbb{R}}dx^\prime H(x^\prime)c_{m}(x^\prime)ln{\frac{|x-L|}{|x-x^\prime|}}
    \label{eq:concentration1}
\end{equation}
Note that the integration constant $C$ does not appear explicitly in this expression and therefore plays no role. We define a dimensionless parameter $\tilde{\epsilon}=\frac{k_{on}L}{D}$ and focus on the regime where $\tilde{\epsilon}>>1$, as it reflects the conditions typically observed in the experiment.

In the limit where ${\tilde \epsilon}$ tends to infinity the integral in Eq.~\eqref{eq:concentration1} vanishes because the concentration $\delta c(x,y=0)$ remains finite as can be seen from Eq.~\eqref{eq:s23}. This imposes that  the solution is $H(x)c_m(x)=0$, so that the concentration of motors close to the filament vanishes. We now proceed by perturbation around this value and calculate $H(x) c_m(x)$ at first order in $1/\tilde \epsilon$. We start from the equation for the Fourier transform of the concentration \eqref{eq:torque2} that we rewrite as
\begin{equation}
\frac{L|q|}{\tilde{\epsilon}}\delta \bar{c}(q) = -\mathcal{F}[H(x)c_m(x)]+\frac{v_0}{2k_{on}} \rho(L) e^{-iqL}  
\end{equation}
At zeroth order in $1/\tilde \epsilon$,  $\delta c(x)=-c_0 H(x)$. Fourier transformation gives $\delta \bar{c}(q)=- \frac{c_0}{iq}(1-e^{-iqL})$ and $\mathcal{F}[H(x)c_m(x)](q)=\frac{c_0}{iq}(1-e^{-iqL})+ \frac{v_0}{2k_{on}} \rho(L)e^{-iqL}$. The value of the concentration of motors in contact with the filament is obtained by inverse Fourier transformation:
\begin{equation}
   H(x) c_m(x) = \frac{c_0 D L}{\pi k_{on}x(L -x)}+ \frac{v_0}{2k_{on}} \rho(L) \delta{[x-L]} \label{cm}  
\end{equation}
Note that the perturbation expansion is singular and that the concentration $c_m(x)$ diverges at $x=0,L$. We will introduce here a cut-off a of the order of the size $d$ of a motor or the radius of the microtubules. The second term in the right hand side of Eq.~\eqref{cm} is due to the flux of motors from the microtubule toward the external space and therefore must be ignored in the calculation of the number of bound motors. 

The concentration of bound motors along the microtubule is given by Eq.~\eqref{eq:boundm}

\begin{equation}
    \rho(x)=\frac{2  c_0 D}{v_0\pi} \log {\frac{x L}{(L -x)d}}
    \label{eq:boundmotor}
\end{equation}

Therefore, the total number of motors on a filament of length $L$ is:
\begin{equation}
    n=\frac{2D c_0L}{\pi v_0}\log(L/d)
    \label{eq:number}
\end{equation}
Three implicit approximations have been made in the calculation of the number of bound motors. We have assumed that the binding rate of motors is large enough that $\tilde{\epsilon} = k_{on}L/D \gg 1$. We also have ignored the excluded volume interactions between motors on the filaments. They become important if the number density of bound motors at the tip of the filaments $\rho(L)$ becomes of the order of the maximum possible density $\rho_m=1/2d$ i.e. if the concentration of free motors is $c_0 \sim \frac{d v_0}{D}$.
Finally, we ignored the detachment of the  motors from the filament. Note also that we calculate here the number of motors in a situation where the microtubule has a vanishing velocity.  

A full discussion of the number of bound motors over the whole parameter range goes beyond the scope of this work. Note however that in the limit where $\tilde{\epsilon}$ is small the concentration $c_m$ is close to $ c_0$ and by integration 
\begin{equation}
n=\frac{k_{on} c_0 L^2}{v_0}
\end{equation}
The number of bound motors indeed vanishes if the binding rate vanishes. 

We performed numerical Brownian dynamics simulations of a single microtubule in the situation studied here and we found good agreement with Eq.~\eqref{eq:number}.
In order to allow for a finite binding rate, we write in the main text $n=\epsilon \frac{k_{on} \rho_0 \hat{c}_0 L^2}{v_0}$, where $\epsilon= \frac{\pi \tilde \epsilon}{2 \log(L/d)}$ if $\tilde\epsilon< 1$ and $\epsilon=1$ if $\tilde \epsilon>1$.

One should note that as in the one-dimensional theory of Ref.~\cite{utzschneider_force_2024}, the number of bound motors is very large when the velocity $v_0$ becomes small. The previous  results are not valid in this limit because the density of attached motors becomes larger than the maximum linear density along the filament $\rho_{max}\sim \mathcal{A}/d$, where $\mathcal{A}$ is a numerical constant that takes into account the geometry of the filament. An estimation of the limit of validity of the theory can be obtained comparing the density $\rho (L)$ to this maximum density. It leads to a minimum duty ratio
$\Re >\Re_{min}\sim \hat{c}_0/\mathcal{A}$. This limits the range of validity of the theory to the smallest motor concentrations. Note however that this is an upper bound because we did not take into account the unbinding rate of the motors that would further decrease $\Re_{min}$ At smaller velocities, the number of bound motors saturates and all the active effects vanish.

\subsection{Calculation of the microtubule velocity}\label{sec:mtvelocity}

As the membrane has a high viscosity, inertia is negligible and we write the force balance condition in the overdamped limit as $f_M+ f_{mt}=0,$ where $f_M=-\xi \Big(n_+ (v_0-v_{mt})+n_-(-v_0-v_{mt})\Big){\p}$ is the total force exerted by the motors on the  microtubule; here $\xi$ is the friction coefficient of a bound motor, $v_{mt}$  the speed of the microtubule, and $n_+$ and $n_-$ are the number of bound motors of types $\pm$, respectively. We consider identical motors with the same gliding speed $v_0$. Moreover, $f_{mt}=-\xi_{mt} v_{mt}\p$ is the drag force exerted on the microtubule where $\xi_{mt}$ is its friction coefficient. Therefore, the speed of the microtubule reads, 
 \begin{equation}
      v_{mt}=-v_0\dfrac{(n_+-n_-)\xi}{\xi_{mt}+(n_++n_-)\xi}\p
\end{equation}
We take here the limit where the friction coefficient of the microtubule is small, so that the speed of the microtubule is

\begin{equation}
      {v_{mt}}=-v_0\dfrac{(n_+-n_-)}{(n_++n_-)}{\p}
     \label{microtubulespeed}
 \end{equation}

\subsection{Calculation of the torque on a microtubule}\label{sec:torque}

In this section, we calculate the torque exerted by bound motors on a microtubule moving on a two-dimensional substrate. 
We consider that the microtubule is made of two parallel filaments of length $L$ at a distance $2a$. For simplicity, we 
suppose that $a$ is equal to the radius $d$ of a molecular motor. The microtubule interacts with a single type of molecular 
motors with a constant concentration gradient $m$ along the vertical direction $c(x,y)=-my+c_0+ \delta c$, where $\delta 
c$ is the excess concentration due to the microtubule. As for the calculation of the number of motors on a filament, we 
impose  external force and torque on the microtubule so that its center of mass and its orientation remain fixed. 

In the limit where the binding rate $k_{on}$ is large, it is a good approximation to assume that there is no motor in the gap between the microtubules and that the flux of motors to each microtubule is approximately half of the value that it would have if there were only one microtubule given by Eq.~\eqref{eq:number}
\begin{equation}
    n_{u,d}=\frac{D({\hat{c}}_0 \rho_0\pm m d)L}{\pi v_0}\log(L/d)
\end{equation}
where we labeled by $u$ the microtubule located at $y=+d$ and by $d$ the microtubule located at $y=-d$.

The torque $\tau$ exerted on a microtubule is given by,
\begin{equation}
    \tau=\int_0^L \Big( \rho_u d f - \rho_d d f \Big)dx,
\end{equation}
where $f$ is the force exerted by a bound motor on the microtubule.

\begin{equation}
    \tau=m L d^2 \xi D\frac{2\log(L/d)}{\pi} 
    \label{eq:torque_expression}
\end{equation}
Note that as the number of motors decreases as $1/v_0$, the torque is independent of the motor velocity $v_0$.
Note also that the assumptions that we made to obtain Eq.~\eqref{eq:torque_expression} are the same as those that we used to calculate the number of motors bound to a microtubule.  As for the total number of motors, we take into account the finite value of the binding rate, $k_{on}$, by writing $\tau=\tilde{\epsilon} m L d^2 \xi D\frac{2\log(L/d)}{\pi}$

\section{Estimation of the parameters\label{parameter_estimation}}

We have constructed the free energy from general principles and not from a microscopic theory. Here we try to estimate the values of the coarse-grained parameters ($\nu$, $\beta$, $\kappa$, and $\zeta$) from the microscopic details or from experiments. We use the estimated values of the parameters in the numerical work.

\subsection{Densities and area fractions}
We start by estimating (from experiments) the microtubule and motor densities. The microtubule concentration is $\phi\approx 1\; \mathrm{MTs}/\mu m^2$ and the concentrations of motors, $\cpm$, in the range  $\sim 1-150 \;\mathrm{motors}/\mu m^2$. The length and diameter of the microtubules are $L \approx \; 5 \mu m$ and $a\approx \; 25 \; nm$, respectively. Moreover, the motors have an effective size of the order of $d\approx 15 \; nm$, estimated from their binding radius \cite{Jiang2019}. For simplicity, we approximate $d =a/2\approx 20 nm = 0.02 \mu m$. Therefore, a microtubule can host roughly $2Ld/(\pi d^2)\approx 160$ bound motors. Furthermore, we get $\rho_0=1/(\pi d^2)\approx 795\; \mu m^{-2}$ using estimated parameters. Therefore, the
area-fraction of microtubules is $\psi = 2 \phi L d \approx 0.2$, and the area fraction of the both motors, $\hat{c}_\pm = \cpm/\rho_0$ varies in the range of $\approx 0.005-0.15$.

\subsection{Physical parameters}

Motors move at a velocity of the order $v_0\approx 1 \;\mu m/s$, and the value of their diffusion coefficient is $D\approx 1.0 \;\mu m^2/s$ from experiments. 
The diffusion coefficient of the microtubules ${D_{mt}}$ is smaller and we estimate the ratio between the two using the the two-dimensional hydrodynamics of the membrane. The diffusion coefficient 
of the membrane is $D= \frac{f k_BT \log[R/d]}{\eta_{2d}}$ where f is a constant, $\eta_{2d}$ is the 2-dimensional 
viscosity and $R$ is a macroscopic length scale which we choose of the order of the period of the pattern $R\sim 10 L$. 
Similarly, the diffusion coefficient of the microtubules is $D_{mt}= \frac{f k_BT \log[R/L]}
{\eta_{2d}}$ leading to $D_{mt} \sim D/3$. The rotational diffusion coefficients of a microtubule 
is of order $D_r \sim D_{mt}/L^2  s^{-1}$. Moreover, from the theoretical calculation of the torque, as 
shown in Eq. \ref{eq:torque_expression}, we get $\nu = \epsilon \frac {2 L d^2 \log{L/d}}{\pi} = 0.007$. However, the values for $\kappa$ and $\zeta$ are not known. We 
study systematically the effect of these two parameters by fixing one of the 
parameters and tuning the other one in our numerical work.

\begin{table}[ht!]
\centering
\begin{tabular}{ll}
\toprule
\textbf{Dimensional parameters} & \textbf{Estimated value of the parameters} \\ 
\midrule
Diffusion constant of the motors $(D)$ & $1.0 \;\mu m^2 / s$ \\
Speed of the bound motors $(v_0)$ & $0.25 \;\mu m / s$ \\
Radius of both motor types  ($d$) & $0.02 \;\mu m$ \\
Concentration of the motors ($c_\pm$) & $1-150 / \mu m^2$ \\
Diffusion constant to the microtubules ($D_{mt})$ & $0.1 \;\mu m^2 / s$ \\
Length of the microtubules and characteristic length ($L$) & $5 \;\mu m$ \\
Rotational diffusion constant of the microtubules $(D_r)$ & $0.1 / s$ \\
Concentration of the microtubules ($\phi$) & $1.0 / \mu m^2$ \\

\bottomrule
\end{tabular}
\label{tableI}
\caption{Dimensional parameters}
\end{table}

\begin{table}[ht!]
\centering
\begin{tabular}{ll}
\toprule
\textbf{Adimensional parameters} & \textbf{Value used in simulations} \\ 
\midrule
Diffusion constant of the motors ($\tilde{D}=\frac{D}{D_r L^2}$) & 0.4 \\
Adimensional speed ($\tilde{v}_0=\frac{v_0}{D_r L}$) &  0.5 \\
Area fraction of the motors varies between($\hat{c}_\pm^0 = c_\pm /\rho_0 )$ & tuned between 0.0065-0.15  \\
Diffusion constant to the microtubules ($\tilde{D}_{mt}=\frac{D_{mt}}{D_r L^2})$ & 0.04 \\
Area fraction of the microtubules ($\psi =  2d L \phi)$ & 0.2 \\
Duty ratio $\Re=\frac{\pi d v_0}{D \log(L/d)}$ & 0.0028 \\
The coefficient $\beta$ & 1.0 \\
Coupling between concentration gradient and polarization($\bar{b}=\frac{\pi D d\rho_0 \nu \Re}{2 v_0 L^3}(=\frac{ d^2}{L^2}))$ & 0.000016 \\
Activity parameter ($\bar a=\frac{2 \epsilon^2 \nu v_0 \hat{c}_0\psi_0\rho_0}{\beta \Re D} (=\frac{4 \epsilon^2 \log^2(L/d)\hat{c}_0 \psi_0 \rho_0 L d}{\beta \pi^2})$ ) & tuned between $1\pm\delta-29.52$ \\
Coefficient of the torque term(${\tilde{\nu}}=\frac{{\tilde v_0}\nu \rho_0}{L})$ & 0.55\\
Bending modulus(${\tilde \kappa= \frac{\pi d\rho_0\kappa\psi_0}{2 \beta L^3}})$ & 0.00001 \\
Gradient energy coefficient(${\tilde \zeta}=\frac{\zeta \rho_0 \hat{c}_0}{L^2})$ & tuned between 0.9 to 2800 \\

\bottomrule
\end{tabular}
\caption{Adimensional parameters}
\label{adimnumbers}
\end{table}


\section{Numerical details and different phases\label{numerical_details}}

The numerical integration of the dynamical equations, Eq.~ \ref{adim-equations}, has been done using a forward finite difference method, available in the authors' github page. We consider a two-dimensional lattice of size $\mathcal{L} \times \mathcal{L}$ with periodic boundary conditions. In all numerical integrations, we start with a homogeneous distribution of both motors and microtubules, with non-zero average value of area fractions  $\hat{c}_\pm^0$ and $\psi_0$, respectively. The average polarisation vanishes in the initial condition. Therefore, we can write the initial distribution for all fields as follows, $\hat{c}_\pm^0=\hat{c}_0+\delta\hat{c}_0$, $\psi=\psi_0+\delta\psi_0$ and $\p=\textbf{0}+\delta\p$, where $\delta\psi_0$ and $\delta\p$ are chosen from a uniform distribution between $[-0.025:0.025]$ with mean $0.0$, and $\delta\hat{c}_0$ are chosen from a uniform distribution between $[-0.005:0.005]$ with mean $0.0$ as well. From linear stability analysis, the peak of the eigenvalue indicates the fastest-growing wavelength. Therefore, we choose a grid size for numerical integration such that the fastest growing wavelength contains at least 10 grid points. An important point to note, is that for each parameter set, we take care of this constraint by choosing appropriate grid size $dx$. For the snapshots in the upper panel in Fig. 3.a in the main text, we have chosen a grid size $dx=0.2$ and $dt=0.0005$ for $\bar{\zeta}=2800$ and $\bar{a}=1.5$, and  $dx=0.015$ and $dt=0.000075$ for $\bar{\zeta}=1.87$ and $\bar{a}=29.52$ for the snapshots in the bottom panel. 

For the phase diagram in the $(\tilde{\zeta},\bar{a})$ plane, as shown in Fig.~3.b of the main text, we have tuned the strength of the interfacial tension $\zeta$, and the area fraction of the two motors are tuned between $0.025$ and $0.15$. We fixed the average area fraction of the microtubules at $\psi_0=0.2$. We choose the same speed and the diffusion constant for both motors and we tabulated the values all the parameters used in the simulations in \ref{adimnumbers} .The coefficient $\alpha$ controls local nonlinear saturation of the polarity field and fixes its steady-state magnitude. We work in the strong-saturation limit to suppress amplitude fluctuations and focus on orientation-driven pattern formation. Its value increases if we increase the area fraction of the motors or decrease the strength of the interfacial tension; $\alpha$ is ranging from $10$ to $1000$ for our numerical integrations. 

\begin{figure}[h]
\centering
\includegraphics[width=0.7\textwidth]{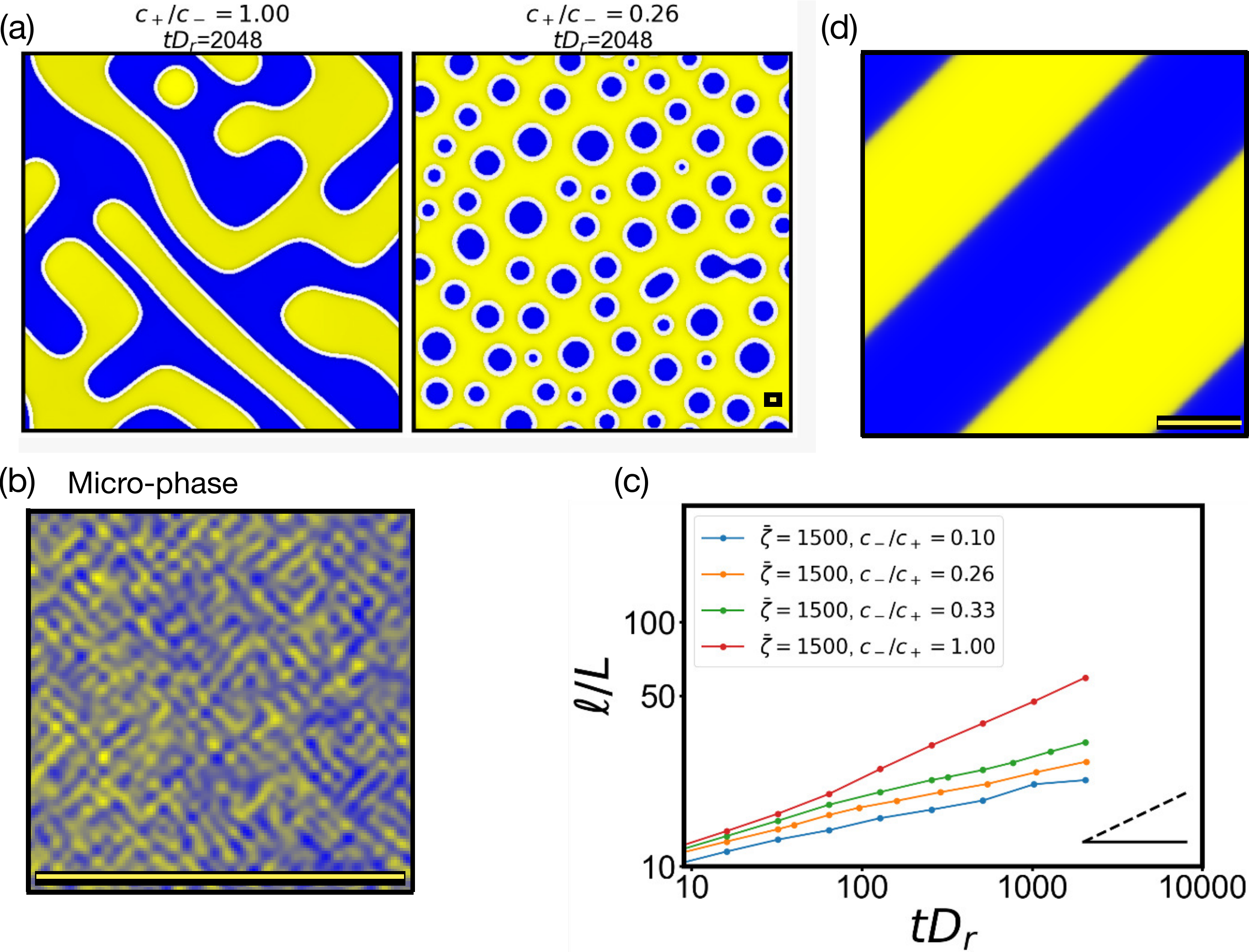}
\caption{(a) Snapshots at time $tD_r= 2048$ for different ratios of two opposite polarity motors. (b) Snapshot of a micro-phase at time $tD_r= 100$. 
(c) Variation of the average domain length $\ell / L$ with time for different ratios of the motor concentrations. We have given the values of $\bar{\zeta}$ in the captions. Dashed and solid lines refer to the slope $1/3$ and $0$, respectively.
(d) Steady state band formation is shown for symmetric case ($c_+/c_-=1.0$) for $\bar{a}=29.52, \bar{\zeta}=0.03$, and $\bar{\kappa}=1250$, and $dx=0.015, dt=0.00005$. The scale bars are $L$, notice the difference in scale bars in different sub figures.}
\label{fig:SI}
\end{figure}

Furthermore, the characteristic domain size for both meso and macro phases grows as, $\ell (t) = t^{\xi}$, where $t$ and $\xi$ refer to the time and growth exponent, respectively.  The domain size $\ell (t)$ at time $t$ is defined as,  $\ell (t)=\frac{2\pi}{\langle {q} \rangle}_t$, where the first moment of $S({q},t)$ is defined as, $\langle {q} \rangle_t=\frac{\int {q} S({q},t) dq}{\int S({q},t)dq}$. The structure factor $S({q},t)$ quantifies spatial correlations in the area fraction field for one of the motors. The structure factor is calculated in simuations by taking the Fast Fourier Transform of one motor density and averaging over all directions for a given magnitude of the wavevector. To understand the coarsening behavior, we have shown the interpolated exponent $\xi$ using color plot in Fig.~3.b in the main text. For macrophase separation, the value of the exponent is $0.33$ and it is $\approx 0.0$ when there is meso/arrested phase separation. As mentioned in the main text, our linear stability analysis predicts type I instability for $\bar{a}>1$. We note that for small enough $\bar{\zeta}$ values there is a presence of meso/arrested phase separation, where $\bar{\xi} \approx 0.0$, as shown in blue in the color plot in Fig. 3 b main text. However, for sufficiently high $\bar{\zeta}$, the system always shows macrophase separation indicating type I instability, as shown in red in the color plot. Furthermore, we have shown the time evolution of the domains for both macro and meso phases in Fig. 3 a, upper and lower panel, respectively. The snapshots are shown at  $D_rt = 64 , 8000 $ and $28000$. It is clear that the domain size saturates for the snapshots shown in the lower panel, whereas the size of the domains grows with time in the upper panel in Fig.~3a of the main text (please note the scale bars in both panels). We note the formation of steady state bands  for higher values of the bending modulus $\tilde{\kappa}$ and very small values of the gradient energy coefficient $\tilde{\zeta}$, as shown in  Fig. \ref{fig:SI}d.  The linear stability analysis also predicts type II / micro phase separation for small values of $\bar{\zeta}$ and $\bar{a} <1$, which is shown with the pink color with solid triangle symbols in Fig.~2a in the main text. We show the snapshot for micro phase separation in \ref{SI_FIG} in Fig. \ref{fig:SI}b. In this case, we have used $dx=0.0025$ and $dt=0.000001$. We have used $\bar{a}=0.98$ and $\bar{\zeta}=0.125$. In this regime the phase separation is dominated by the torque on the microtubules exerted by the motors and not by the active transport. Therefore, the characteristic size of the domains is very small, and can be even less than the length of one microtubule (please note the scale bar in the snapshot).

We have also done a numerical study for  asymmetric concentrations of the two motors, and
we found droplet phases of one motor type in a continuous phase of the other motor. The snapshots at time $D_rt= 2048$ for different concentration ratios of the two motors are shown in Fig. \ref{fig:SI}a. We have not performed the linear stability analysis in the asymmetric case, but in the numerical study we have used the same grid size $dx$ and a smaller time step $dt$ as in the symmetric case where the sum of average concentrations of the two motors are identical. Furthermore, using the same approach, as discussed in the previous paragraph, we have calculated the characteristic domain size. The variation of the characteristic domain size with time is shown for different ratios of the motor concentrations, in  Fig. \ref{fig:SI}c. Interestingly, we note that for any given value $\bar{\zeta}$ and $\bar{a}$, coarsening of the domains slows down with the asymmetry of the motor concentrations, which suggests spatially stable aster formation for asymmetric motor concentrations. 

\section{Supplementary figures\label{SI_FIG}}
This section contains one supplementary figure on pattern formation in the numerical work for different values of the parameters from those of the main text and in the section. \ref{numerical_details}, and for asymmetric mixtures of different concentrations.

\newpage
\bibliographystyle{apsrev4-1} 
\bibliography{biblio}